\documentclass{article}
\usepackage{amsmath}
\usepackage{amsfonts}

\usepackage{amsmath,amssymb}
\usepackage{graphicx}
\usepackage{color}

\newtheorem{theorem}{Theorem}
\newtheorem{lemma}{Lemma}

\newtheorem{definition}{Definition}

\newtheorem{remark}{Remark}

\newcommand{\ls}[1]
    {\dimen0=\fontdimen6\the\font\lineskip=#1\dimen0
     \advance\lineskip.5\fontdimen5\the\font
     \advance\lineskip-\dimen0
     \lineskiplimit=0.9\lineskip
     \baselineskip=\lineskip
     \advance\baselineskip\dimen0
     \normallineskip\lineskip\normallineskiplimit\lineskiplimit
     \normalbaselineskip\baselineskip
     \ignorespaces}

\begin{document}

\bibliographystyle{abbrv}

\title{The detailed autocorrelation distribution and 2-adic complexity of a classes of binary sequences with almost optimal autocorrelation}
\author{Yuhua Sun$^{1,2,3}$, Qiang Wang$^{2}$, Tongjiang Yan$^{1,3}$\\
$^1$ College of Sciences,
China University of Petroleum,\\
Qingdao 266555,
Shandong, China\\
$^2$ School of Mathematics and Statistics,\\
Carleton University,
Ottawa ,
Ontario,K1S 5B6, Canada\\
$^3$ Key Laboratory of Network Security and Cryptology,\\
Fujian Normal University,
Fuzhou, Fujian 350117,
China\\
Email: sunyuhua\_1@163.com;\ wang@math.carleton.ca;\ yantoji@163.com\\
}

 \maketitle
\footnotetext[1] {The work is supported by Shandong Provincial Natural Science Foundation of China(No. ZR2014FQ005, The Fundamental Research Funds for the Central Universities(No. 15CX02065A, No. 15CX08011A, No. 15CX02056A, No. 16CX02013A, No. 16CX02009A), Fujian Provincial Key Laboratory of Network Security and Cryptology Research Fund(Fujian Normal University)(No.15002).
}

\thispagestyle{plain} \setcounter{page}{1}

\begin{abstract}
Pseudo-random sequences with good statistical property, such as low autocorrelation, high linear complexity and large 2-adic complexity, have been used in designing reliable stream ciphers.
In this paper, we obtain the exact autocorrelation distribution of a class of sequence with three-level autocorrelation  and analyze the 2-adic complexity of this sequence.
Our results show that the 2-adic complexity of the sequence  is at least $(N+1)-\mathrm{log}_2(N+1)$ and that in many cases it is maximal, which is large enough to resist the attack of the rational approximation algorithm (RAA) for feedback with carry shift registers (FCSRs).

\end{abstract}

\ls{1.5}
\section{INTRODUCTION}\label{section 1}
Pseudo-random sequences with good statistical property are widely used as basic blocks for constructing stream ciphers.   Any key stream generators could be implemented by both linear feedback shift registers (LFSRs) and feedback with carry shift registers (FCSRs). However, after the Berlekamp-Massey algorithm (BMA) for LFSRs \cite{Massey} and the rational approximation algorithm for FCSRs \cite{Andrew Klapper} were presented, linear complexity and 2-adic complexity of the key stream sequence have been regarded as the critical security criteria and required to be no less than one half of the period. Autocorrelation is another critical statistical measure of the key stream sequence.
Although the linear complexity of many classes of sequences have been obtained (See \cite{Ding Cunsheng}-\cite{Edemskiy}), there are only a handful research papers that focus on 2-adic complexity.  For example,  in 1997, Klapper has pointed out that an $m$-sequence with prime period has maximal 2-adic complexity \cite{Andrew Klapper}. In 2010, Tian and Qi showed that the 2-adic complexity of all the binary $m$-sequences is maximal \cite{Tian Tian}. Afterwards, Xiong et al. \cite{Xiong Hai} presented a new method of circulant matrices to compute the 2-adic complexities of binary sequences.

Several recent results show that the 2-adic complexity of a sequence possesses a close relationship with its another critical statistical property (i.e., autocorrelation).  In \cite{Xiong Hai},  Xiong et al  showed that all the known sequences with ideal 2-level autocorrelation have maximum 2-adic complexity. Moreover, in \cite{Ding-H-L},    Ding et al  proved that the 2-adic complexities of Legendre sequences and Ding-Helleseth-Lam sequences  
 with optimal autocorrelation are also maximal. Then, using the same method as that in \cite{Xiong Hai}, Xiong et al. \cite{Xiong Hai-2} pointed out that two other classes of sequences based on interleaved structure have also maximal 2-adic complexity. One of these two classes of sequences was constructed by Tang and Ding \cite{Tang-Ding}, which has  optimal autocorrelation, the other was constructed by Zhou et al \cite{Zhou}, which is optimal with respect to the Tang-Fan-Matsufuji bound \cite{Tang-Fan bound}. Recently, Hu \cite{Hu Honggang} presented a simpler method to obtain the results of Xiong et al. \cite{Xiong Hai}, using detailed autocorrelation values.

In \cite{Cai Ying}, Cai and Ding gave a generic construction of a large class of sequences with almost optimal autocorrelation, using almost difference sets. Then Wang \cite{Wang Qi} and Sun et al \cite{Sun Yuhua} proved that most of these sequences have high linear complexity. Meanwhile, Sun et al \cite{Sun Yuhua} generalized Cai and Ding's construction using $d$-form function with difference-balanced property and  obtained  more sequences with almost optimal autocorrelation in this way. In this paper, motivated by Hu's method \cite{Hu Honggang}, we determine the exact autocorrelation distribution and obtain a lower bound on the 2-adic complexity of these sequences. Our result shows that the low bound for this class of sequences with period $N$ is at least $(N+1)-\mathrm{log}_2(N+1)$ and that in many cases it is maximal, which is large enough to resist against the rational approximation algorithm (RAA) attack for feedback with carry shift registers (FCSRs).

The rest of this paper is organized as follows. Some neccesary definitions, notations,  and previous results are introduced in Section 2. The exact autocorrelation distribution of  a  class of almost optimal autocorrelation sequences that  generalized from Cai and Ding  \cite{Cai Ying} by Sun et al \cite{Sun Yuhua}  is given in Section 3. In Section 4, the lower bounds on the 2-adic complexities of these sequences will be presented.  Finally we  summarize our results and give some remarks in Section 5.

\section{Preliminaries}\label{section 2}

Let $N$ be a positive integer and $s=(s_{0},s_{1},\cdots,s_{N-1})$ a binary sequence of period $N$. Let $S(x)=\sum\limits_{i=0}^{N-1}s_{i}x^i\in \mathbb{Z} [x]$. Then we write
\begin{equation}
\frac{S(2)}{2^N-1}=\frac{\sum\limits_{i=0}^{N-1}s_{i}2^i}{2^N-1}=\frac{p}{q},\ 0\leq p\leq q,\ \mathrm{gcd}(p,q)=1.\label{2-adic complexity}
\end{equation}
The 2-adic complexity $\Phi_{2}(s)$ of the sequence $s$ is the integer $\lfloor\mathrm{log}_2q\rfloor$, i.e.,
\begin{equation}
\Phi_{2}(s)=\left\lfloor\mathrm{log}_2\frac{2^N-1}{\mathrm{gcd}(2^N-1,S(2))}\right\rfloor,\label{2-adic calculation}
\end{equation}
 where $\lfloor x\rfloor$ is the greatest integer that is less than or equal to $x$.

Let $C=\{0\leq i\leq N-1: s_{i}=1\}$ be the support of $s$. Then $s$  is called the characteristic sequence of $C$. The autocorrelation of $s$ is defined by
\begin{equation}
AC(\tau)=\sum\limits_{i=0}^{N-1}(-1)^{s_i+s_{i+\tau}},\ \tau=0,1,2,\cdots,N-1,\label{AC}
\end{equation}
where $\tau\in \mathbb{Z}_N$. It is well known that $AC(\tau)\equiv N\pmod 4$ for all $\tau\in Z_N$. Moreover, it  can be computed by
\begin{equation}
AC(\tau)=N-4(|C|-d_{C}(\tau)),\label{equation 1}
\end{equation}
where $d_{C}(\tau)$ is the difference function of the support $C$ such that  $\tau+C=\{\tau+i \mid i \in C\}$ and
\begin{equation}
d_{C}(\tau)=|C\cap(\tau+C)|. \label{differ function}
\end{equation}

\begin{definition}\label{definition 1}
Let $(\mathrm{G},+)$ be a cyclic group with $N$ elements and $C$ be  a $k$-element subset of $\mathrm{G}$ . Supposing the variable $\tau$ ranges over all the nonzero elements of $\mathrm{G}$.  If $d_{C}(\tau)$ always takes on the value $\lambda$, then $C$ is called a $(N,k,\lambda)$ cyclic difference set of $\mathrm{G}$; if $d_{C}(\tau)$ takes on $\lambda$ altogether $t$ times and $\lambda+1$ altogether $N-1-t$ times, then $C$ is called a $(N,k,\lambda,t)$ cyclic almost difference set (CADS) in $\mathrm{G}$.
\end{definition}
According to Eq. (\ref{equation 1}), when the support $C$  of a sequence $s$  is a $(N,k,\lambda,t)$ cyclic almost difference set, the autocorrelation of $s$ is
\begin{equation}
AC(\tau)=\left\{
\begin{array}{lll}
N,\ \ \ \ \ \ \ \ \ \ \ \ \ \ \ \ \ \ \ \ \ \ \mathrm{for}\ 1\ \mathrm{ time },\\
N-4(k-\lambda),\ \ \ \ \ \ \ \mathrm{for}\ t\ \mathrm{ times },\\
N-4(k-\lambda-1),\ \ \mathrm{ for\ } N-1-t\ \mathrm{ times }.
\end{array}
\right.\label{CADS}
\end{equation}
Therefore, under the assumption $N\equiv3\pmod 4)$,  a sequence $s$ of period $N$ has almost optimal autocorrelation if and only if $AC(\tau)=-1$ or $3$ for all $\tau \not\equiv0\pmod N$ (see \cite{Cai Ying}).

Let $q$ be a power of a prime and $n$ a positive integer.
\begin{definition}\label{def d-form}  A function $f$  from $\mathbb{F}_{q^{n}}$ to $\mathbb{F}_{q}$ is called a $d$-form function on $\mathbb{F}_{q^{n}}$ over $\mathbb{F}_{q}$ if
$f(xy)=y^{d}f(x)$ for any $x\in \mathbb{F}_{q^{n}}$ and $y\in \mathbb{F}_{q}$.
\end{definition}
\begin{definition} A function from $\mathbb{F}_{q^{n}}$ to $\mathbb{F}_{q}$ is said to be balanced if the element 0 appears one less time than each nonzero element in $\mathbb{F}_{q}$ in the list $f(\alpha^{0}),\ f(\alpha^{1}),\cdots,f(\alpha^{q^{n}-2})$, where $\alpha$ is a primitive element of $\mathbb{F}_{q^{n}}$.
\end{definition}
\begin{definition}\label{D-B-F} Let $f(x)$ be a $d$-form function on $\mathbb{F}_{q^{n}}$ over $\mathbb{F}_{q}$ and $\mathrm{gcd}(d,q^n-1)=1$. Then the function $f(x)$ is called difference-balanced if $f(xz)-f(x)$ is balanced for any $z\in \mathbb{F}_{q^{n}}\setminus\{1\}$.
\end{definition}

Earlier, Sun et al  \cite{Sun Yuhua} extended Cai and Ding's construction  \cite{Cai Ying}  to obtain the following almost difference sets.

\begin{lemma}\label{lemma construction of ADS}
\cite{Sun Yuhua} Let $m$ be a positive integer, $\alpha$ a primitive element of the finite field $\mathbb{F}_{2^{2m}}$ and $f(x)$ a $d$-form function from $\mathbb{F}_{2^{2m}}$
to $\mathbb{F}_{2^{m}}$ with difference-balanced property. Suppose that $C_1^{\prime}$ is any $(2^m-1,2^{m-1}-1,2^{m-2}-1)$ difference set in $(\mathbb{Z}_{2^m-1},+)$. Define $C_{1}=\{(2^m+1)i \mid i\in C_{1}^{\prime}\}$, $C_{2}=\{i\in \mathbb{Z}_{2^{2m}-1} \mid f(\alpha^{i})=1\}$, $C=C_{1}+C_{2}=\{(c_{1}+c_{2}) \pmod{2^{2m}-1} \mid c_{1}\in C_{1}, c_{2}\in C_{2}\}$.
Then $C$ is a $(2^{2m}-1,2^{2m-1}-2^m,2^{2m-2}-2^m,2^{m}-2)$ almost difference set in $(\mathbb{Z}_{2^{2m}-1},+)$. Furthermore, the characteristic sequence of the set $C$ has the out-of-phase autocorrelation values $\{-1,3\}$ only.
\end{lemma}

In the sequel, we also need the following number theoretical results.

\begin{definition}\label{pseudoprime} A composite number $n$ is called a 2-pseudoprime if $2^{n-1}\equiv1\pmod n$.
\end{definition}

For exmple,  both $341=11\cdot31$ and $561=3\cdot11\cdot17$ are 2-pseudoprimes.

\begin{lemma}\label{Pseudoprime thm}\cite{Ramanujachary}
If $n$ is a 2-pseudoprime, then $2^n-1$ is a 2-pseudoprime. Therefore, there are infinitely many 2-pseudoprimes.
\end{lemma}

\section{Detailed autocorrelation distribution of sequences generalized from Cai and Ding by Sun et al.}\label{section 3}

In this section, we  derive the exact autocorrelation distribution of  the sequence $s$ constructed in  Lemma \ref{lemma construction of ADS}.  Previously, we know that the autocorrelation of this sequence is almost optimal, however, it is not  good enough to help us determine the lower bound on its 2-adic complexity.  In order to achieve our goal, we use  Eq. (\ref{CADS}) and Lemma \ref{lemma construction of ADS}  to find out the exact autocorrelation distribution of $s$.

\begin{lemma}\label{property of D-B-F}
Let $f(x)$ be a $d$-form function on $\mathbb{F}_{q^n}$ over $\mathbb{F}_{q}$ with difference-balanced property. Define $H_{a}=\{x\in \mathbb{F}_{q^n}^{\ast}\mid f(x)=a,\ a\in \mathbb{F}_{q}^{\ast}\}$. Then, for a primitive element $\beta$ of $\mathbb{F}_{q}$ and $i\in\{1,2,\cdots,q-2\}$, we must have $\beta^ix\notin H_{a}$ for any  $x\in H_a$ and $|H_{a}|=q^{n-1}$.
\end{lemma}
$\mathbf{Proof.}$ By the definition of $d$-form function, $f(\beta^ix)=\beta^{id}f(x)$ for any $x\in \mathrm{F}_{q^n}$. If $x\in H_a$ and $\beta^ix\in H_a$, we get $\beta^{id}=1$, which is impossible since $\mathrm{gcd}(d,q-1)=1$ and $i\in\{1,2,\cdots,q-2\}$. Additionally, the difference-balanced property guarantees $|H_{a}|=q^{n-1}$.\ \ \ \ \ \ \ \ \ \ \ \ \ \ \ \ \ \ \ \ \ \ \ \ \ \ \ \ \ \ \ \ \ \ \ \ \ \ \ \ \ \ \ \ \ \ \ \ \ \ \ \ \ \ \ \ \ \ \ \ \ \ \ \ \ \

\begin{lemma}\label{relation difference sets}
Let all the symbols be the same as those in Lemma \ref{lemma construction of ADS}. Suppose that $\tau=(2^m+1)\tau_1$, where $\tau_1\in\{1,2,\cdots,2^m-2\}$. Then
$$
d_C(\tau)=|C\cap(C+\tau)|=2^{2m-2}-2^m.
$$
\end{lemma}
$\mathbf{Proof.}$
Let $c_1=(2^m+1)i$ with a fixed $i\in C_1^{\prime}$ such that $i+\tau_1\in C_1^{\prime}$.  Then, for any $c_2\in C_2$, we can see that $c=c_1+c_2\in C$ and $c+\tau=c_1+c_2+\tau=(2^m+1)(i+\tau_1)+c_2\in C$.  Conversely,  for any $c\in C$, the pair $(c_{1},c_{2})$ such that $c= c_1 +c_2$ with $c_1\in C_1$ and $c_2\in C_2$  is unique. Moreover, there exists exactly one $i\in C_1^{\prime}$ such that $c_1=(2^m+1)i$
and    $c+\tau=(2^m+1)(i+\tau_1)+c_2 \in C$. Indeed, by the first conclusion of Lemma \ref{property of D-B-F}, $c_2+(2^m+1)k\notin C_2$ for any $c_2\in C_2$ and any $k\in\{1,2,\cdots,2^m-2\}$. Therefore,  $c=c_{1}+c_{2}\in C$ and $c+\tau\in C$ if and only if $(2^m+1)(i+\tau_1)\in C_1$, i.e., $i+\tau_1\in C_1^{\prime}$.

Hence,
$$d_C(\tau)=|C\cap(C+\tau)|=|C_2|\cdot|C_1^{\prime}\cap(C_1^{\prime}+\tau_1)|=|C_2|\cdot d_{C_{1}^{\prime}}.$$
By the assumption  that $C_{1}^{\prime}$ is a $(2^m-1,2^{m-1}-1,2^{m-2}-1)$ difference set in $(\mathbb{Z}_{2^m-1},+)$. Then $d_{C_{1}^{\prime}}=2^{m-2}-1$. Furthermore, $|C_2|=2^m$ by Lemma \ref{property of D-B-F}. The result follows.
\ \ \ \ \ \ \ \ \ \ \ \ \ \ \ \  \ \ \ \ \ \ \ \ \ \ \ \ \ \ \ \ \ \ \ \ \ \ \ \  \ \ \ \ \ \ \ \ \ \ \ \  \ \ \ \ \ \ \

\begin{theorem}\label{thm3}
Let $m$ be a positive integer, $\alpha$ a primitive element of $\mathbb{F}_{2^{2m}}$ and $f(x)$ a $d$-form function from $\mathbb{F}_{2^{2m}}$
to $\mathbb{F}_{2^{m}}$ with difference-balanced property. Suppose that $C_1^{\prime}$ is any $(2^m-1,2^{m-1}-1,2^{m-2}-1)$ difference set in $(\mathbb{Z}_{2^m-1},+)$. Define $C_{1}=\{(2^m+1) i \mid i\in C_{1}^{\prime}\}$, $C_{2}=\{i\in \mathbb{Z}_{2^{2m}-1} \mid f(\alpha^{i})=1\}$, $C=C_{1}+C_{2}=\{(c_{1}+c_{2})\ \pmod{2^{2m}-1}\mid c_{1}\in C_{1},c_{2}\in C_{2}\}$.
Then the exact autocorrelation distribution of the characteristic sequence $s$ of $C$ is given by
$$
AC(\tau)=\left\{
\begin{array}{ll}
-1,\  \tau\in\{(2^m+1)\tau_1\mid \tau_1=1,2,\cdots,2^m-2\},\\
3,\ \ \ 1\leq\tau\leq2^{2m}-2,\ \mathrm{but}\ \tau\notin\{(2^m+1)\tau_1 \mid \tau_1=1,2,\cdots,2^m-2\}.
\end{array}
\right.
$$
\end{theorem}
$\mathbf{Proof.}$ First of all, from the parameters of the almost difference set $C$ in Lemma \ref{lemma construction of ADS}, we know that $|C|=2^{2m-1}-2^m$ and that there are $2^m-2$ $\tau$'s such that the autocorrelation $AC(\tau)=-1$. Secondly, by Lemma \ref{relation difference sets} we have $d_C(\tau)=|C\cap(C+\tau)|=2^{2m-2}-2^m$ for $\tau\in \{(2^m+1)\tau_1 \mid \tau_1=1,2,\cdots, 2^m-2\}$. Then, using Eq. (\ref{equation 1}), we get $AC(\tau)=-1$ for $\tau\in \{(2^m+1)\tau_1 \mid \tau_1=1,2,\cdots, 2^m-2\}$. Note that the size of the set $\{(2^m+1)\tau_1|\tau_1=1,2,\cdots, 2^m-2\}$ is exactly $2^m-2$. Therefore the proof is complete.

\section{Lower bounds on the 2-adic complexity of sequences generalized from Cai and Ding by Sun et al.} \label{section 4}
Recall that the sequence $s$ in Lemma \ref{lemma construction of ADS} has the period $N=2^{2m}-1$.  In the following we  let $S(x)=\sum\limits_{i=0}^{N-1}s_ix^i$ and $T(x)=\sum\limits_{i=0}^{N-1}(-1)^{s_i}x^i\in \mathbb{Z}[x]$.

\begin{lemma}\label{method}
Let $S(x)=\sum\limits_{i=0}^{N-1}s_ix^i$ and $T(x)=\sum\limits_{i=0}^{N-1}(-1)^{s_i}x^i\in \mathbb{Z}[x]$. Then
\begin{equation}
-2S(x)T(x^{-1})\equiv N+\sum\limits_{\tau=1}^{N-1}AC(\tau)x^{\tau}-T(x^{-1})\left(\sum\limits_{i=0}^{N-1}x^i\right)\pmod{x^N-1}.\label{finatrans}
\end{equation}
\end{lemma}
$\mathrm{proof.}$
 According to the definition of $T(x)$, we have
\begin{eqnarray}
T(x)T(x^{-1})\nonumber
&\equiv&\left(
\sum\limits_{i=0}^{N-1}(-1)^{s_i}x^{i}\right)\left(\sum\limits_{j=0}^{N-1}(-1)^{s_j}x^{-j}\right)\pmod{x^N-1}\nonumber\\
&\equiv&\sum\limits_{i=0}^{N-1}\sum\limits_{j=0}^{N-1}(-1)^{s_i+s_j}x^{i-j} \pmod{x^N-1}\nonumber\\
&\equiv&N+\sum\limits_{\tau=1}^{N-1}\sum\limits_{j=0}^{N-1}(-1)^{s_{j+\tau}+s_j}x^{\tau}\pmod{x^N-1}\nonumber\\
&\equiv&N+\sum\limits_{\tau=1}^{N-1}AC(\tau)x^{\tau}\pmod{x^N-1}.\label{AC2C}
\end{eqnarray}
Furthermore, we have
\begin{equation}\label{999}
T(x)=\sum\limits_{i=0}^{N-1}(-1)^{s_{i}}x^i=\sum\limits_{i=0}^{N-1}(1-2s_{i})x^i=\sum\limits_{i=0}^{N-1}x^i-2S(x).
\end{equation}
Combining Eqs. (\ref{AC2C})$-$(\ref{999}), we obtain the result.
$\Box$

Employing Lemma \ref{method} and the detailed autocorrelation distribution of $s$, we can obtain the following.
\begin{lemma}\label{extend}
Let $m$ be a positive integer, $N=2^{2m}-1$, and $s$ be the binary sequence with almost optimal autocorrelation in Lemma \ref{lemma construction of ADS}. Then
\begin{equation}
S(2)T(2^{-1}) \equiv -2\left((2^{2m-2}-\frac{2^N-1}{2^{2^m+1}-1}\right)\pmod{2^N-1}).
\end{equation}
\end{lemma}
$\mathrm{proof.}$ Recall that $C_{1}=\{(2^m+1)\tau_1|\tau_1=1,2,\cdots,2^m-2\}$ in Lemma \ref{method}. In convenience, we denote $\mathbf{Z}_N^{\star}=\{1,2,\cdots,N-1\}$. Substituting the autocorrelation in Theorem \ref{thm3} into Eq. (\ref{finatrans}) in Lemma \ref{method}, we can see
\begin{eqnarray}
-2S(x)T(x^{-1})&\equiv& N+\sum\limits_{\tau=1}^{N-1}AC(\tau)x^{\tau}-T(x^{-1})\left(\sum\limits_{i=0}^{N-1}x^i\right)\pmod{x^N-1}\nonumber\\
&\equiv&N+\sum\limits_{\tau\in \mathbf{Z}_N^{\star}\setminus C_{1}} 3\cdot x^{\tau}+\sum\limits_{\tau\in C_{1}}(-1)\cdot x^{\tau}\nonumber\\
&&-T(x^{-1})\left(\sum\limits_{i=0}^{N-1}x^i\right)\pmod{x^N-1}\nonumber\\
&\equiv&N+\sum\limits_{\tau\in \mathbf{Z}_N^{\star}} 3\cdot x^{\tau}+\sum\limits_{\tau\in  C_{1}} (-3)\cdot x^{\tau}+\sum\limits_{\tau\in C_{1}}(-1)\cdot x^{\tau}\nonumber\\
&&-T(x^{-1})\left(\sum\limits_{i=0}^{N-1}x^i\right)\pmod{x^N-1}\nonumber\\
&\equiv&N+\sum\limits_{\tau\in \mathbf{Z}_N^{\star}} 3\cdot x^{\tau}+\sum\limits_{\tau\in  C_{1}} (-4)\cdot x^{\tau}\\
&&-T(x^{-1})\left(\sum\limits_{i=0}^{N-1}x^i\right)\pmod{x^N-1}\nonumber\\
&\equiv&N+\sum\limits_{\tau=1}^{N-1} 3\cdot x^{\tau}+\sum\limits_{\tau_1=1}^{2^m-2}(-4)\cdot x^{(2^m+1)\tau_1} \nonumber\\
&&-T(x^{-1})\left(\sum\limits_{i=0}^{N-1}x^i\right)\pmod{x^N-1}\nonumber\\
&\equiv&N-3\left(1-\sum\limits_{\tau=0}^{N-1}x^{\tau}\right)+\left(4-4\cdot\sum\limits_{\tau_1=0}^{2^m-2}x^{(2^m+1)\tau_1}\right)\nonumber\\
&&-T(x^{-1})\left(\sum\limits_{i=0}^{N-1}x^i\right)\pmod{x^N-1}\nonumber\\
&\equiv&N+1-\frac{4(x^N-1)}{x^{2^m+1}-1}-(1+T(x^{-1}))\left(\sum\limits_{i=0}^{N-1}x^i\right)\pmod{x^N-1}.\nonumber
\end{eqnarray}
Then, substituting 2 for $x$ we have
\begin{eqnarray}
-2S(2)T(2^{-1})&\equiv&N+1-\frac{4(2^N-1)}{2^{2^m+1}-1}\pmod{2^N-1})\nonumber\\
&\equiv&4\left(2^{2m-2}-\frac{2^N-1}{2^{2^m+1}-1}\right)\pmod{2^N-1}).\nonumber
\end{eqnarray}
The result follows.
$\Box$

In order to give the lower bound on the 2-adic complexity, we also need the following simple result in number theory whose proof is omitted here.

\begin{lemma}\label{numtheory}
Let $n^{\prime},\ m^{\prime}$ be any positive integer. Then
\[\displaystyle\frac{2^{n^{\prime}m^{\prime}}-1}{2^{m^{\prime}}-1}\equiv n^{\prime}\pmod{2^{m^{\prime}}-1}.\]
\end{lemma}

Now we give a general lower bound on the 2-adic complexity for the sequence $s$ constructed in Lemma~ \ref{lemma construction of ADS}.

\begin{theorem}\label{generalbound}
Let $m$ be a positive integer, $\alpha$ a primitive element of $\mathbb{F}_{2^{2m}}$ and $f(x)$ a $d$-form function from $\mathbb{F}_{2^{2m}}$
to $\mathbb{F}_{2^{m}}$ with difference-balanced property. Suppose that $C_1^{\prime}$ is any $(2^m-1,2^{m-1}-1,2^{m-2}-1)$ difference set in $(\mathbb{Z}_{2^m-1},+)$, $C_{1}=\{(2^m+1)i|i\in C_{1}^{\prime}\}$, $C_{2}=\{i\in \mathbb{Z}_{2^{2m}-1}|f(\alpha^{i})=1\}$  and $C=C_{1}+C_{2}=\{(c_{1}+c_{2})\pmod(2^{2m}-1)|c_{1}\in C_{1},c_{2}\in C_{2}\}$. Denote $s$ to be the characteristic sequence of $C$. Then the 2-adic complexity $\Phi_{2}(s)$ of $s$ is bounded by
\begin{equation}
\Phi_{2}(s)\geq (N+1)-\mathrm{log}_2(N+1).\label{genebound}
\end{equation}
\end{theorem}
$\mathbf{Proof.}$ Above all, from the known conditions, we know that $m\geq2$.
It is easy to see that
$$
\mathrm{gcd}\left(S(2),2^N-1\right)|\mathrm{gcd}\left(S(2)T(2^{-1}),2^N-1\right)
$$
and that the product
$$\mathrm{gcd}\left(S(2)T(2^{-1}),\frac{2^N-1}{2^{2^m+1}-1}\right)\mathrm{gcd}\left(S(2)T(2^{-1}),2^{2^m+1}-1)\right)$$
is divided by $\mathrm{gcd}\left(S(2)T(2^{-1}),2^N-1\right)$.
Then we get
\begin{eqnarray}
\mathrm{gcd}\left(S(2),2^N-1\right)|\left(\mathrm{gcd}\left(S(2)T(2^{-1}),\frac{2^N-1}{2^{2^m+1}-1}\right) \right.\nonumber\\
\left.\mathrm{gcd}\left(S(2)T(2^{-1}),2^{2^m+1}-1\right)\right).\label{expand}
\end{eqnarray}
By Lemma \ref{extend}, we have
\begin{eqnarray*}
S(2)T(2^{-1})&\equiv&-2(2^{2m-2}-\frac{2^N-1}{2^{2^m+1}-1})\pmod{2^N-1}\nonumber\\
&\equiv&-2^{2m-1}\pmod {\frac{2^N-1}{2^{2^m+1}-1}}.\nonumber
\end{eqnarray*}
Then we have
\begin{equation}
\mathrm{gcd}(S(2)T(2^{-1}),\frac{2^N-1}{2^{2^m+1}-1})=\mathrm{gcd}(2^{2m-1}, \frac{2^N-1}{2^{2^m+1}-1})=1. \label{seconddiscussion}
\end{equation}
Upper bound of $\mathrm{gcd}\left(S(2)T(2^{-1}),2^{2^m+1}-1\right)$ is considered in the following. Note that $N=2^{2m}-1=(2^m-1)(2^m+1)$. By Lemma \ref{numtheory}, we get
$$\frac{2^N-1}{2^{2^m+1}-1}\equiv2^m-1\pmod{2^{2^m+1}-1}.$$
Then, by Lemma \ref{extend} again, we have
\begin{eqnarray*}
S(2)T(2^{-1})&\equiv&-2\left(2^{2m-2}-\frac{2^N-1}{2^{2^m+1}-1}\right)\pmod{2^{N}-1}\nonumber\\
&\equiv&-2\left(2^{2m-2}-2^m+1\right)\pmod{2^{2^m+1}-1}\nonumber
\end{eqnarray*}
Accordingly, we know that
\begin{equation}
\mathrm{gcd}\left(S(2)T(2^{-1}),2^{2^m+1}-1\right)=
\mathrm{gcd}\left(2^{2m-2}-2^{m}+1, 2^{2^m+1}-1\right).\label{firstdiscussion}
\end{equation}
Note that
$$2^{2m-2}-2^{m}+1<2^{2m-2}+1<2^{2^m+1}-1\ \ (m\geq2).$$
Therefore we obtain an upper bound of $\mathrm{gcd}(S(2)T(2^{-1}),2^{2^m+1}-1)$
\begin{equation}
\mathrm{gcd}(S(2)T(2^{-1}),2^{2^m+1}-1)\leq 2^{2m-2}-2^{m}+1<2^{2m-2}-1.\label{firstneq}
\end{equation}

Combining Eqs. (\ref{expand}),(\ref{seconddiscussion}) and (\ref{firstneq}), we have
$$
\mathrm{gcd}(S(2),2^N-1)\leq 2^{2(m-1)}-1.
$$
Therefore, by Eq. (\ref{2-adic calculation}), the 2-adic complexity $\Phi_2(s)$ of $s$ is bounded by
\begin{eqnarray*}
\Phi_2(s)&=&\lfloor\mathrm{log}_2\frac{2^N-1}{\mathrm{gcd}(S(2),2^N-1)}\rfloor\geq (N-1)-2(m-1)\nonumber\\
&=&N-2m+1=(N+1)-\mathrm{log}_2(N+1).\nonumber
\end{eqnarray*}
\ \ \ \ \ \ \ \ \ \ \ \ \ \ \ \ \ \ \ \ \ \ \ \ \ \ \ \ \ \ \ \ \ \ \ \ \ \ \ \ \ \ \ \ \ \ \ \ \ \ \ \ \ \ \ \ \ \ \ \ \ \ \ \ \ \ \ \ \ \ \ \ \ \ \ \ \ \ \ \ \ \ \ \ \ \ \ \ \ \ \ \ \ \ \ \ \ \ \ \ \ \ \ \ \ \ \ \ \ \ $\Box$
\begin{remark}
From the result of Theorem \ref{generalbound}, it is easy to test that the lower bound in Eq.(\ref{genebound}) is larger than $\frac{N}{2}$ for any positive integer $m$. Hence, the 2-adic complexity of $s$ is large enough to resist RAA. In fact, in many cases, the lower bound can be maximal. In the following, we will discuss these cases.
\end{remark}

\begin{lemma}\label{finallemma2}
Let $m-1$ be a prime or a 2-pseudoprime. Then we have
\begin{equation}
\mathrm{gcd}\left(S(2)T(2^{-1}), 2^{2^m+1}-1\right)=\left\{
\begin{array}{lllll}
2^5-1,\ \mathrm{if}\ m-1\equiv5\pmod{20},\nonumber\\
1,\ \ \ \ \ \ \ \ \mathrm{otherwise}.
\end{array}
\right.\label{2-pseudorandom}
\end{equation}
\end{lemma}
$\mathbf{Proof.}$ Note that $2^{2m-2}-2^{m}+1=(2^{m-1}-1)^2$. Then, by Eq. (\ref{firstdiscussion}), we know that
\begin{eqnarray}
\mathrm{gcd}\left(S(2)T(2^{-1}), 2^{2^m+1}-1\right)=\mathrm{gcd}\left((2^{m-1}-1)^2,2^{2^m+1}-1\right).\label{2-pseu-gcd-0}
\end{eqnarray}

Next, we will determine the value of $\mathrm{gcd}\left(2^{m-1}-1,2^{2^m+1}-1\right)$ in several steps.

Firstly, it is easy to see that
\begin{equation}
\mathrm{gcd}\left(2^{m-1}-1,2^{2^m+1}-1\right)=2^{\mathrm{gcd}(m-1,2^m+1)}-1.\label{2-pseu-gcd-1}
\end{equation}
 Since $m-1$ is a prime or a 2-pseudoprime, then we have $m-1|2^{m-2}-1$, which implies that $2^m+1=4(2^{m-2}-1)+5\equiv5\bmod m-1$.
 Therefore, we have $\mathrm{gcd}(2^m+1,m-1)=\mathrm{gcd}(m-1,5)$. By Eq. (\ref{2-pseu-gcd-1}), we know that
 \begin{equation}
\mathrm{gcd}\left(2^{m-1}-1,2^{2^m+1}-1\right)=\left\{
\begin{array}{ll}
2^5-1,\ \mathrm{if}\ m-1\equiv0\pmod{5},\\
1,\ \ \ \ \ \ \ \ \mathrm{otherwise}.
\end{array}
\right.\label{2-pseu-gcd-2}
\end{equation}

Secondly, we will prove $m-1\equiv5\pmod{10}$ if $5|m-1$, i.e., $m-1$ is not even if $5|m-1$. Otherwise, if $m-1$ is even, i.e., $m$ is odd, then $m\equiv1\pmod{4}$ or $m\equiv3\pmod{4}$, which implies that
$2^m\equiv2\pmod{5}$ if $m\equiv1\pmod{4}$ and $2^m\equiv3\pmod{5}$ if $m\equiv3\pmod{4}$. But since $m-1$ is a prime or 2-pseudoprime, we have $m-1|2^{m-2}-1$. Further, since $5|m-1$, then $5|2^{m-2}-1$ and $2^m=4(2^{m-2}-1)+4\equiv4\pmod{5}$, a contradiction to $2^m\equiv2\pmod{5}$ or $2^m\equiv3\pmod{5}$. Thus, we obtain that
\begin{equation}
\mathrm{gcd}(2^{m-1}-1, 2^{2^m+1}-1)=\left\{
\begin{array}{lllll}
2^5-1,\ \mathrm{if}\ m-1\equiv5\bmod{10},\nonumber\\
1,\ \ \ \ \ \ \ \ \mathrm{otherwise}.
\end{array}
\right.\label{2-pseu-gcd-4}
\end{equation}

Thirdly, we will prove $m-1\equiv5\pmod{20}$ if $m-1\equiv5\pmod{10}$, i.e., $m-1\not\equiv15\pmod{20}$ if $m-1\equiv5\pmod{10}$. Otherwise, if $m-1\equiv15\pmod{20}$, then we have $m-1\equiv3\pmod{4}$ or $m\equiv0\pmod{4}$, which implies $2^m\equiv1\pmod{5}$, a contradiction to the above $2^m=4(2^{m-2}-1)+4\equiv4\pmod{5}$. Thus we have
\begin{equation}
\mathrm{gcd}(2^{m-1}-1, 2^{2^m+1}-1)=\left\{
\begin{array}{lllll}
2^5-1,\ \mathrm{if}\ m-1\equiv5\pmod{20},\nonumber\\
1,\ \ \ \ \ \ \ \ \mathrm{otherwise}.
\end{array}
\right.\label{2-pseu-gcd-5}
\end{equation}

Now, we will prove that $2^{2^m+1}-1$ is not divided by $(2^5-1)^2$ if $2^5-1|2^{2^m+1}-1$. Otherwise, it is easy to see that $2^5-1|2^{2^m+1}-1$, i.e., $5|2^m+1$. Then, we have $2^{2^m+1}-1=(2^5-1)\times\frac{2^{2^m+1}-1}{2^5-1}$, which implies that $2^5-1|\frac{2^{2^m+1}-1}{2^5-1}$. By Lemma \ref{numtheory}, we have $\frac{2^{2^m+1}-1}{2^5-1}\equiv\frac{2^m+1}{5}\bmod2^5-1$. Then we can get $2^5-1|2^m+1$. Therefore, we have $2^5-1|2^{2m}-1$, i.e., $5|2m$, which implies $5|m$, a contradiction to the fact $5|m-1$. Thus, by Eqs. (\ref{2-pseu-gcd-0}) and (\ref{2-pseu-gcd-5}), we know that Eq. (\ref{2-pseudorandom}) holds. The desired result follows.
$\Box$

\begin{theorem}\label{2-adicofads}
Let $m$ be a positive integer, $N=2^{2m}-1$, $\alpha$ a primitive element of $\mathbb{F}_{2^{2m}}$ and $f(x)$ a $d$-form function from $\mathbb{F}_{2^{2m}}$ to $\mathbb{F}_{2^{m}}$ with difference-balanced property. Suppose that $C_1^{\prime}$ is any $(2^m-1,2^{m-1}-1,2^{m-2}-1)$ difference set in $(\mathbb{Z}_{2^m-1},+)$. Define $C_{1}=\{(2^m+1)i|i\in C_{1}^{\prime}\}$, $C_{2}=\{i\in \mathbb{Z}_{2^{2m}-1}|f(\alpha^{i})=1\}$, $C=C_{1}+C_{2}=\{(c_{1}+c_{2})\pmod(2^{2m}-1)|c_{1}\in C_{1},c_{2}\in C_{2}\}$. Let $s$ be the sequence whose support set is $C$. Then the 2-adic complexity $\Phi_{2}(s)$ of $s$ is bounded by
\begin{eqnarray*}
\Phi(s)\geq\left\{
\begin{array}{lllll}
N-1,\ \ \ \ \ \ \ \ \ \ \ \ \ \ \ \ \ \ \ \mathrm{if}\ m-1\ is\ a\ prime\ or\\
\ \ \ \ \ \ \ \ \ \ \ \ \ \ \ \ \ \ \ \ \ \ \ \ \ \ \ a\ 2-pseudoprime\ but\ m-1\not\equiv5\pmod{20},\\
N-6,\ \ \ \ \ \ \ \ \ \ \ \ \ \ \ \ \ \ \ \mathrm{if}\ m-1\ is\ a\ prime\ or\\
\ \ \ \ \ \ \ \ \ \ \ \ \ \ \ \ \ \ \ \ \ \ \ \ \ \ \ a\ 2-pseudoprime\ and\ m-1\equiv5\pmod{20},\\
(N+1)-\mathrm{log}_2(N+1),\ \mathrm{otherwise}.
\end{array}
\right.\nonumber
\end{eqnarray*}
\end{theorem}
$\mathbf{Proof.}$ From Eq. (\ref{2-adic calculation}), the 2-adic complexity of $s$ satisfies
\begin{eqnarray*}
\Phi(s)&=\lfloor\mathrm{log}_2\frac{2^N-1}{\mathrm{gcd}(S(2),2^N-1)}\rfloor\geq \lfloor\mathrm{log}_2\frac{2^N-1}{\mathrm{gcd}(S(2)T(2^{-1}),2^N-1)}\rfloor\nonumber\\
&\geq\lfloor\mathrm{log}_2\frac{2^N-1}{\mathrm{gcd}(S(2)T(2^{-1}),2^{2^m+1}-1)\mathrm{gcd}(S(2)T(2^{-1}),\frac{2^N-1}{2^{2^m+1}-1})}\rfloor\nonumber
\end{eqnarray*}
The rest of the proof is from Lemma \ref{finallemma2} and the discussion in the proof of Theorem \ref{generalbound}. \ \ \ \ \ \ \ \ \ \ \ \ \ \ \ \ \ \ \ \ \ \ \ \ \ \ \ \ \ \ \ \ \ \ \ \ \ \ \ \ \ \ \ \ \ \ \ \ \ \ \ \ \ \  \ \ \ \ \ \ \ \ \ \ \ \ \ \ \ \ \ \ \ \ \ \
\begin{remark}
From  Theorem \ref{2-adicofads}, the 2-adic complexity of the sequences with almost optimal autocorrelation  is large enough to resist the analysis of RAA.
\end{remark}

\section{Summary and concluding remarks}
In this paper, we first gave the detailed autocorrelation distribution of the sequence with almost optimal autocorrelation generalized from Cai and Ding by Sun et al.. Then, using the the detailed autocorrelation distribution and combining the method of Hu and some number theory, we present the lower bounds on the 2-adic complexity of these sequences in the general case and some special cases respectively.
Our results show that the 2-adic complexity is at least $(N+1)-\mathrm{log}_2(N+1)$ and that in many cases it is maximal, which is obviously large enough to resist RAA of FCSR.

\end{document}